\documentclass[prb,superscriptaddress,a4paper,aps,twocolumn,showpacs]{revtex4}
\usepackage{graphicx,bm,color}

\def\--{\negthinspace - \negthinspace}

\begin{document}

\title{Ground-state path integral Monte Carlo
simulations of positive ions in $^4$He clusters: 
bubbles or snowballs?
}

\author{
  Stefano Paolini
}
\email{
  stefano.paolini@pd.infn.it
}
\affiliation{
   -- Dipartimento di Fisica ''G. Galilei'', Universita' di Padova, 
  via Marzolo 8, I-35131 Padova, Italy
}
\affiliation{
  CNR-INFM--DEMOCRITOS National Simulation Center, Trieste, Italy
}
\author{
  Francesco Ancilotto
}
\email{
  francesco.ancilotto@pd.infn.it
}
\affiliation{
   -- Dipartimento di Fisica ''G. Galilei'', Universita' di Padova, 
  via Marzolo 8, I-35131 Padova, Italy
}
\affiliation{
  CNR-INFM--DEMOCRITOS National Simulation Center, Trieste, Italy
}

\author{
  Flavio Toigo
}
\email{
  flavio.toigo@pd.infn.it
}
\affiliation{
   -- Dipartimento di Fisica ''G. Galilei'', Universita' di Padova, 
  via Marzolo 8, I-35131 Padova, Italy
}
\affiliation{
  CNR-INFM--DEMOCRITOS National Simulation Center, Trieste, Italy
}

\date{
  \today
}

\begin{abstract}

The local order around alkali (Li$^+$ and Na$^+$) and alkaline-earth 
(Be$^+$, Mg$^+$ and Ca$^+$)
ions in $^4$He clusters has been studied using ground-state path integral Monte 
Carlo calculations. We apply a criterion based on multipole dynamical correlations 
to discriminate between solid-like versus liquid-like behavior of the $^4$He shells 
coating the ions.   
As it was earlier suggested by experimental measurements in bulk $^4$He, our findings 
indicate that Be$^+$ produces a solid-like (``snowball'') structure, similarly to 
alkali ions and in contrast to the more liquid-like $^4$He structure embedding heavier 
alkaline-earth ions.

\end{abstract}

                                                                                                             

\maketitle

\section{Introduction}

In past decades, the implantation of ions in liquid helium has often been 
 used to probe the bulk properties of the superfluid 
state.~\cite{fetter} The attention has also focused on the details 
of the structure of the helium solvent around the ions.~\cite{atkins,bachman,flavio-1978} 
As a consequence of electrostriction, 
a strong increase in the helium density with respect to its bulk value
is expected in the surroundings of the impurity site.~\cite{atkins,bachman}
Within a crude picture, ions in liquid helium can be divided in two 
categories, on the basis of the different local structure they create 
in the solvent. 
Strongly attractive ions tend to form
a \emph{solid-like} structure, 
the so-called \emph{snowball}, characterized  by  
a very inhomogeneous helium density in the surroundings of the impurity.
Alkali ions are believed to belong to this category.
In contrast, 
singly charged alkaline-earth cations, due to their
larger radii and weaker interaction with the He atoms, are expected 
to produce a cavity surrounded by compressed, 
less structured,  and most likely \emph{not solidified}, helium 
(\emph{bubble} in the following).~\cite{bachman} 

Several investigations of the ion-helium 
structure have been based on measurements 
of ions mobilities in bulk helium.~\cite{glaberson-1975,foerste-1997}        
In particular, the experimental results of Ref.~\onlinecite{foerste-1997}
have shown that the temperature dependence of 
the mobility of Be$^+$ is much more similar to that of alkali ions 
(K$^+$,Rb$^+$ and Cs$^+$)~\cite{glaberson-1975} and He$^+$,~\cite{foerste-1997}  
than to that of the other alkaline-earth ones 
(Mg$^+$,Ca$^+$,Ba$^+$,Sr$^+$).~\cite{foerste-1997}
It was thus suggested that Be$^+$ forms a snowball, 
similarly to He$^+$ and alkali ions, and at variance 
with heavier alkaline-earth ions, where ``bubble" states 
are believed to develop instead.~\cite{foerste-1997} 
The recent discovery that positive ions can be 
captured in helium nanodroplets~\cite{stienkemeier-2003} 
promises an extension to charged particles of 
experimental techniques which proved 
extraordinarily fruitful for neutral 
species.~\cite{toennies-2004,lehmann-2006,barranco}

Quantum many-body simulation methods allow for a microscopic 
description of such systems,~\cite{buzzacchi-2001,na+nakayama} 
as well as for the investigation of solid-like behavior 
in quantum clusters.~\cite{baroni-moroni-chemphyschem}
Using quantum Monte Carlo (QMC) simulations, within the shadow wave function 
approach, 
different positive ions have been studied both in bulk  
$^4$He,~\cite{buzzacchi-2001,rossi-2004} and $^4$He clusters,~\cite{rossi-2004}
with the aim of computing  
the effective masses of charged impurities moving in liquid $^4$He 
 and of establishing, at the same time, the surrounding local structure of the solvent 
(for a calculation of the effective masses within the hydrodynamical model see 
Ref.~\onlinecite{lehmann-2002}).

The distinction between solid- and liquid-like behavior 
of $^4$He around atomic/molecular impurities is a subtle issue 
which requires a careful examination of the dynamical evolution 
of the system. 
Within reptation quantum Monte Carlo (RQMC),~\cite{RQMC-cornell,RQMC-PRL} 
the use of suitable (imaginary) time-correlation functions  
has been exploited to propose a dynamical criterion for discriminating between 
freezing versus quantum melting in small para-hydrogen 
clusters.~\cite{baroni-moroni-chemphyschem} 
In the present paper, we resort to the same criterion
to investigate the structural order in $^4$He clusters doped with
alkali and alkaline-earth positive ions, simulated by a more standard 
ground-state path integral (PIGS) Monte Carlo algorithm.~\cite{schmidt-PIGS} 
We take Li$^+$ and Na$^+$ 
as representatives of alkali metals, while 
among the alkaline-earths, we consider Be$^+$, Mg$^+$ and Ca$^+$.
Lithium and sodium -doped $^4$He clusters have been chosen because
the strong interaction of these ions with the $^4$He solvent
makes them prototypes of 
snowball structures, whose expected solid-like signature can provide 
a useful benchmark in the study of the other ions, where a clear-cut
distinction between bubbles and snowballs might not be easy to 
be resolved. 
As a paradigmatic case of liquid helium-impurity structures, 
we also simulated a small $^4$He cluster seeded with a carbon monoxide 
molecule.
We considered ion impurities
in clusters rather than in bulk helium in order to avoid the possibility
of finding fictitious structures imposed by the symmetry
of the periodically repeated simulation box.
Moreover, the local $^4$He structure around the ion
in large enough clusters should be very similar to that developed 
in bulk $^4$He.~\cite{rossi-2004}

We show that the criterion proposed in Ref.~\onlinecite{baroni-moroni-chemphyschem}
is applicable also to larger systems than those investigated in that 
paper, and that it is able to discriminate between a crystalline 
or a liquid behavior of the impurity-He structure. 
In qualitative agreement with experimental data,~\cite{foerste-1997}
our simulations show in particular that the $^4$He structure around Be$^+$ is much more 
similar to that found around alkali ions, where a snowball structure is found, 
than to the one around heavier alkaline-earth ions, where $^4$He exhibits  
a more liquid-like behavior.

Our paper is organized as follows. In Sec.~II we briefly review our theoretical framework. 
Section~III~A contains our results for very small $^4$He clusters (seeded with Li$^+$ and CO),  
that have been considered for testing our computational strategy. 
In Sec.~III~B we present simulations for 
several ions solvated in larger $^4$He clusters, and we discuss the implications of these results on 
the main issue we are addressing in this paper (i.e. the local structure of the $^4$He 
solvent around Be$^+$ in 
comparison with the findings for alkali ions and other alkaline-earth ones). Finally, 
Section~IV contains our conclusions.

\section{Theory}
\label{sec:theory}

Our clusters are described by a realistic Hamiltonian, $\widehat{H}$, in which 
the $N$ $^4$He atoms and the ionic impurity are treated as point particles interacting 
via pair potentials. When the CO molecule is used as dopant, we neglect its internal degrees 
of freedom and we treat it as a rigid linear 
rotor with only translational and rotational degrees of freedom. 
The interactions are derived from accurate quantum chemistry calculations, either 
by an interpolation of tabulated values, as for Li$^+$-He and 
Ca$^+$-He,~\cite{pot-Li,pot-Ca} or by parametrized analytical expression, as for 
He-He,~\cite{potenziali-He-He} He-CO,~\cite{Heijmen} and for the remaining  
He-ions pairs.\cite{pot-Na,pot-Be-Mg}  
The size of the ion-doped clusters varies up to a maximum of 70 $^4$He atoms, whereas, 
in the case of CO, we considered a smaller cluster, with only 
fifteen particles (CO@He$_{15}$), 
approximatively corresponding to one complete solvation shell.~\cite{CO-RQMC} 

We simulate the system using the ground-state path integral Monte Carlo scheme,~\cite{schmidt-PIGS} 
where the imaginary-time evolution operator, $e^{-\beta \widehat{H}}$, is exploited 
to improve systematically a suitable trial function $\Psi_T$, by projecting it out onto 
the state $|\Psi_{\beta}\rangle=e^{-\beta\widehat{H}}|\Psi_T\rangle$. The latter 
converges to the exact ground state, $\Psi_0$, for $\beta\to\infty$.   
By Trotter slicing the imaginary-time evolution into $m$ time steps of length $\epsilon=\beta/m$, 
and by choosing an appropriate definite positive short-time approximation 
for the imaginary-time propagator, 
$G({\bf{R}}^\prime,\bf{R};\epsilon)\approx\langle\bf{R}^\prime|e^{-\epsilon\widehat{H}}|{\bf{R}}
\rangle$ between the configurations ${\bf{R}}$ (a vector which describes the coordinates of all 
$^4$He atoms in the cluster, in addition to those of the impurity) and  ${\bf{R}}^\prime$, 
expectation values of quantum operators, $\widehat{O}$,
on the state $|\Psi_{\beta}\rangle$,
can be expressed in a path-integral form:

\begin{equation}
\langle\widehat{O}\rangle_{\beta}=\frac{\langle\Psi_{\beta}|\widehat{O}|\Psi_{\beta}\rangle}
{\langle\Psi_{\beta}|\Psi_{\beta}\rangle}\approx\int dX O({\bf{R}}_{m}) P(X), 
\label{O_beta}
\end{equation}

\noindent
where $P(X)=\Psi_T({\bf{R}}_0)\prod_{i=1}^{2m} 
G({\bf{R}}_{i},{\bf{R}}_{i-1};\epsilon)\Psi_T({\bf{R}}_{2m})$, 
$X=\{{\bf{R}}_0,\cdots,{\bf{R}}_{2m}\}$ is a path 
of configurations of the system, and $\widehat{O}$ 
is diagonal in the coordinate representation.
By sampling the probability $P(X)$ via a generalized Metropolis algorithm, 
and accumulating the values assumed by the operator $\widehat{O}$ over the sampled paths, 
the desired expectation value can be accessed within a known statistical error and without 
any mixed estimate nor population-control bias.

For the relatively small CO@He$_{15}$, it is sufficient to use the primitive approximation~\cite{ceperley-RMP} 
for the imaginary-time propagator $G$, and then to sample the probability $P(X)$ by the RQMC algorithm.~\cite{RQMC-cornell,RQMC-PRL}
In the case of ionic impurities, we resorted to the pair approximation for $G$,~\cite{ceperley-RMP}
employing the bisection-multilevel sampling procedure.~\cite{ceperley-RMP}
The dependence of $\langle\widehat{O}\rangle_{\beta}$ on $\beta$ and $\epsilon$ needs 
to be checked, in order to extrapolate the result to the exact value.  
Our projection times $\beta$ varied between $0.1 \rm K^{-1}$ and 
$0.7 \rm K^{-1}$, while we used time steps $\epsilon$ ranging between $0.0025 \rm~K^{-1}$ 
and $0.0125 \rm~K^{-1}$ for clusters seeded with ions, and equal to 
 $0.001 \rm~K^{-1}$ for CO@He$_{15}$. These values of $\beta$ and $\epsilon$ have been 
empirically found to ensure the stability of our results.

We used a Jastrow-type trial wave function:
\begin{equation}
\Psi_T=\textrm{exp}\Big[ -\sum_{i=1}^N u_1(r_i) -\sum_{i<j}^N u_2(r_{ij})\Big],
\label{Psi_T}
\end{equation}

\noindent 
where $r_i$ and $r_{ij}$ indicate, respectively, the distances between the $i$-th $^4$He atom 
and the ion, and that between the $i$-th and the $j$-th $^4$He atoms. 
The radial functions $u_1$ and $u_2$ have been optimized independently for each one of the 
simulated systems, by minimizing the cluster variational energy with respect to 
a small number (10-12) of adjustable parameters.
For CO@He$_{15}$, the one-body term $u_1$ also depends on the angle between the molecular 
axis and the vector position of the $^4$He atom with respect to the molecule center of mass 
(details can be found in Ref.~\onlinecite{CO-RQMC}). 

Within the path integral scheme, $\widehat{O}$ can be either a static operator 
(such as the Hamiltonian itself) or a time-dependent one, such as those needed 
to calculate imaginary-time correlation functions, 
$\widehat{A}\textrm{exp}(-\tau\widehat{H})\widehat{B}$.
Their evaluation requires an extended time range, $2\beta+\tau$, but it is nevertheless 
straightforward: $c_{AB}(\tau)=\langle\Psi_{\beta}|\widehat{A}(\tau_0)
\widehat{B}(\tau_0+\tau)|\Psi_{\beta}\rangle$ is estimated as $\langle \frac{1}{n_i}\sum_i 
A({\bf{R}}_i)B({\bf{R}}_{i+k})\rangle$, 
where $k=\tau/\epsilon$, 
$\langle\cdot\rangle$ indicates a statistical average over the sampled paths, 
and the $n_i$ values of $i$ span the range $\beta/\epsilon<i<2m-2\beta/\epsilon-k$.
The results that will be shown in the following Sections have been obtained 
by varying $\tau$ within  $\rm 1~K^{-1}$. 
Notice that time-translation invariance implies that these correlations are independent of 
$\tau_0$.

As it was proposed in Ref.~\onlinecite{baroni-moroni-chemphyschem}, 
imaginary-time correlations can be exploited to unveil either a solid- or 
a liquid-like behavior in quantum clusters, 
by monitoring the persistence of geometrical signatures 
of a solid-like structure. 
The criterion of Ref.~\onlinecite{baroni-moroni-chemphyschem}  
is based on the definition of the rotationally invariant time-correlation functions 

\begin{equation}
c_L(\tau)=\frac{\sum_M\langle Q_{LM}^*(\tau_0)\overline{Q}_{LM}(\tau_0+\tau)\rangle}
{\sum_M \langle Q_{LM}^*(\tau_0)Q_{LM}(\tau_0)\rangle},
\label{multipole-correlations}
\end{equation}

\noindent
of the multipole moments, 

\begin{equation}
Q_{LM}=\sqrt{4\pi/(2L+1)}\int d{\bf{r}} \rho({\bf{r}}) 
r^L Y^*_{LM}(\theta,\phi),
\label{multipole-def}
\end{equation}

\noindent
of the cluster mass distribution $\rho({\bf{r}})$, around its center of mass.
In Eq.~\ref{multipole-correlations}, $Q_{LM}(\tau_0)$ is the multipole moment of the system 
at (imaginary) time $\tau_0$ and $\overline{Q}_{LM}(\tau_0+\tau)$ indicates the multipole calculated 
after the configuration at time $\tau_0+\tau$ has been rotated back in order 
to minimize the particles diffusion between the two time instants. 

\begin{figure}[t]
  \hbox to \hsize{\hfill
    \includegraphics[width=75mm]{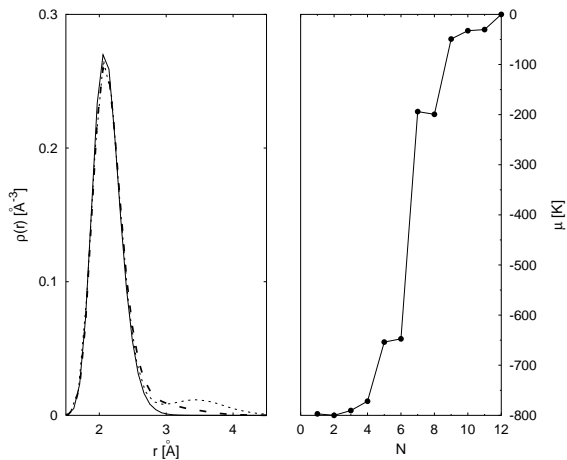}
    \hfill
  }
  \caption{
    Left panel: Radial density distribution of $^4$He atoms around the 
    Li$^+$ ion in Li$^+$@He$_N$ for $N=8$ (solid line), $N=9$ (long dashed line), 
    and $N=10$ (short dashed line).
    Right panel: $^4$He chemical potential $\mu_N=E_N-E_{N-1}$ as a function of the 
    cluster size, $N$, in Li$^+$@He$_N$, whose total energy is $E_N$.  
     }
  \label{fig:rho-mu-Li+He_8}
\end{figure}

For a rigid system $c_L(\tau)$ is independent of time, whereas it decays to zero at large 
$\tau$ if the system undergoes random fluctuations. A large value of $c_L(\tau)$ at large 
times is thus a distinctive mark of the persistence of the shape (``rigidity'') of the system. 
This criterion has been originally applied to small clusters of para-hydrogen molecules 
(both pure an doped with a CO molecule), for cluster sizes around twelve molecules, at which a first 
solvation shell is completed.~\cite{baroni-moroni-chemphyschem}
As a preliminary step in the study of the correlation functions $c_L(\tau)$, we 
evaluate the static multipole correlation function, 
$q_L=\sum_M\langle Q^*_{LM}Q_{LM}\rangle$, which provides a 
rotationally invariant characterization of 
a rigid body.~\cite{baroni-moroni-chemphyschem}
This quantity is not suitable for discriminating the cases where 
a given multipole vanishes on average but $q_L$ is different from zero because 
of fluctuations, from those where a non-vanishing $q_L$ value is due 
to the average shape of the system. However,  
non-vanishing $q_L$ values allow to select a set of $L$'s 
among which to search for  slow-decaying time-correlations $c_L(\tau)$. 

\begin{figure}[t]
  \hbox to \hsize{\hfill
    \includegraphics[width=75mm]{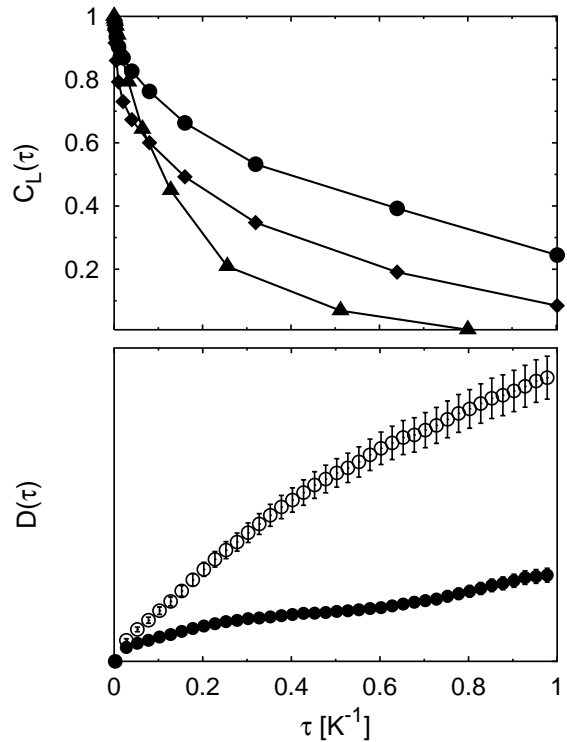}
    \hfill
  }
  \caption{
    Upper panel: Multipole time-correlation functions $c_L(\tau)$ of 
    Li$^+$@He$_8$ ($L=4$: diamonds, $L=5$: filled circles) and 
    CO@He$_{15}$ ($L=5$: triangles). 
    Lower panel: Diffusion coefficient of the $^4$He atoms in Li$^+$@He$_8$, 
    as calculated from the original configurations (empty circles) and 
    from configurations rotated back as described in the text
    (filled circles). 
     }
  \label{fig:dinamica-Li+He_8}
\end{figure}

\section{RESULTS AND DISCUSSION}

\subsection{Testing the approach on small clusters}

As a test case, we first discuss the results for a Li$^+$ ion in 
small $^4$He clusters (Li$^+$@He$_N$). The left panel of Fig.~\ref{fig:rho-mu-Li+He_8} 
displays the radial distribution functions of $^4$He atoms around the impurity ion, 
for cluster sizes in the range $N=8-10$. For $N\geq9$ the magnitude of the maximum 
of these functions stays nearly constant while the tail of the distribution 
extends to larger distances from the dopant. Correspondingly, the chemical 
potential rises steeply from $N=8$ to $N=9$. These findings indicate that 
for $N=8$ a first solvation shell has been completed. We notice 
that an even more marked jump in the chemical potential occurs at $N=6$, 
thus suggesting a great stability of Li$^+$@He$_6$. In particular 
some of the multipole imaginary-time correlation functions of Li$^+$@He$_6$ show 
an extremely slow decay in imaginary time, which is indicative of the rigidity of the 
system (see below). However, since we aim at revealing the existence of solid-like signatures within 
the first shell of larger clusters, we will not discuss any longer the case of 
Li$^+$@He$_6$, focusing instead on the solvent structure in a cluster of eight $^4$He 
atoms, i.e. one representing a fully developed first solvation shell.

\begin{figure}[t]
  \hbox to \hsize{\hfill
    \includegraphics[width=75mm]{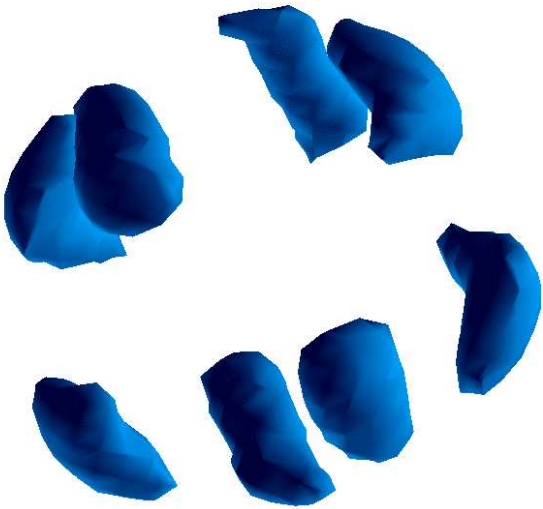}
    \hfill
  }
  \caption{
     He density distribution for Li$^+$@He$_8$. 
     The displayed isosurface contains about $50\%$ of the total number of 
     He atoms. The Li$^+$ ion (not shown) is located at the center of the cluster.
     }
  \label{fig:rho-Li+He_8}
\end{figure}

The upper panel of Fig.~\ref{fig:dinamica-Li+He_8} shows 
the multipole time-correlation functions of Li$^+$@He$_8$ for $L=4,5$ and, 
for comparison, of CO@He$_{15}$ for $L= 5$.  
All the multipoles correlations of CO@He$_{15}$ decay rapidly, 
indicating a liquid-like behavior of the first He shell around the molecule,
and we arbitrarily 
chose to show the $L=5$ curve as representative of the general behavior 
of the $c_L(\tau)$ correlations for any $L$. 
On the other hand, in Li$^+$@He$_8$, all the multipole time-correlations  
with $L<4$ decay rapidly, 
whereas for $L=4$ and, more evidently, for $L=5$, after a short transient 
in which a very steep drop occurs, the correlations decays much more slowly, 
indicating the long persistence of an average geometrical shape of the cluster. 
In the lower panel of Fig.~\ref{fig:dinamica-Li+He_8} we compare 
the diffusion coefficient 
$D(\tau)=\langle\frac{1}{N}\sum_{i=1}^N[\bf{r}_i(\tau+\beta)-\bf{r}_i(\beta)]^2\rangle$ 
of the $^4$He atoms, as calculated from the original 
configurations of each quantum path generated during the simulation, with that obtained 
after rotating backward the configurations at time 
$\tau$ in order to minimize the particles diffusion with respect to the configuration 
at time $\beta$. The high degree of rigidity of the cluster structure 
near the impurity is confirmed by the 
considerable effectiveness of the diffusion minimization procedure.

In order to visualize the intrinsic shape of the $^4$He cluster, we 
have calculated the helium density distribution 
$\overline{\rho}(\bf{r})=\langle\sum_{i=1}^N\delta(\bf{r}-\bf{r}_i)\rangle$
around the ion, by applying the mentioned backward-rotation procedure.  
The contribution to $\overline{\rho}(\bf{r})$ from each quantum path 
sampled by the PIGS simulation is then further rotated to minimize 
the diffusion of the particles centroids of different 
paths. Fig~\ref{fig:rho-Li+He_8} shows the density 
distribution for Li$^+$@He$_8$. The solvent accumulates in eight 
regions roughly arranged in two parallel squares, rotated by $\pi/4$
with respect to each other. 
Indeed, the lowest non-vanishing $q_L$ values for a polyhedron with such a structure 
correspond to $L=4$ and $L=5$. 
The results reported in Figs.~\ref{fig:dinamica-Li+He_8} and~\ref{fig:rho-Li+He_8} 
seem to indicate that Li$^+$@He$_8$ behaves much like a rigid body
rather than as an ion embedded within a liquid shell.

\begin{figure}[t]
  \hbox to \hsize{\hfill
    \includegraphics[width=75mm]{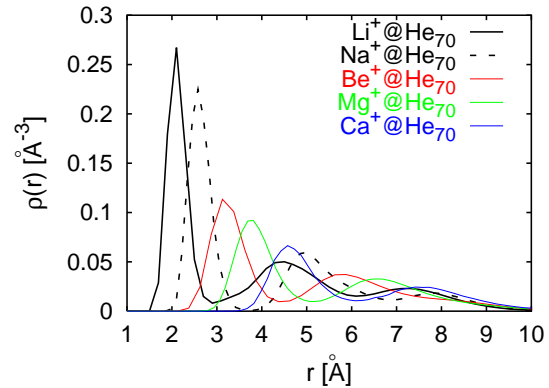}
    \hfill
  }
  \caption{Radial density distribution function of the He atoms in 
   Li$^+$@He$_{70}$ (solid black line),  Na$^+$@He$_{70}$ (dashed line), 
   Be$^+$@He$_{70}$ (red),  Mg$^+$@He$_{70}$ (green),  Ca$^+$@He$_{70}$ 
   (blue).
     }
  \label{fig:radial-density}
\end{figure}

\subsection{Multiple-shell ion-doped $^4$He clusters}

We come now to the main concern of this paper, i.e. to discriminate
between solid-like versus liquid-like structures around helium-solvated ions. 
To this aim we have tried to apply the criterion of Ref.~\onlinecite{baroni-moroni-chemphyschem} 
to the more general case of clusters consisting of more than one 
solvation shell.

Instead of extending the integral in Eq.~\ref{multipole-def} to the entire cluster, 
one may calculate multipole moments $Q_{LM}$ of a region within a suitably chosen distance from the 
center of mass. Thus, the criterion described in the previous Section 
allows to identify a solid-like behavior of that sub-region of the cluster. 
In Fig.~\ref{fig:radial-density} we show the radial solvent density profiles 
for $70$ $^4$He atoms clusters doped with different ions. 
By taking the first minima of the curves as cut-off distances in the integral which 
defines $Q_{LM}$, we accessed the imaginary-time auto-correlation functions 
of the multipoles of the first shell of the corresponding clusters.  
Table I reports the integral of the $^4$He density within the 
first solvation shell for the various ions that we have considered.

\begin{table}[b]
\caption{
Integral of the $^4$He density distribution within the first 
solvation shell of~$^4$He$_{70}$ clusters doped with different ions.
 }
\begin{tabular}{|c|c|c|c|c|}
\hline
\hspace{0.3cm} \textbf{Li$^+$ \hspace{0.3cm}}  
& \hspace{0.3cm} \textbf{Na$^+$ \hspace{0.3cm}}  
& \hspace{0.3cm} \textbf{Be$^+$ \hspace{0.3cm}}  
& \hspace{0.3cm} \textbf{Mg$^+$ \hspace{0.3cm}}  
& \hspace{0.3cm} \textbf{Ca$^+$ \hspace{0.3cm}} \\
\hline
8.24 & 12.02 & 14.55 & 19.02 & 23.18\\
\hline
\end{tabular}
\end{table}

\subsubsection{Alkali ions doped $^4$He clusters}

\begin{figure}[t]
  \hbox to \hsize{\hfill
    \includegraphics[width=75mm]{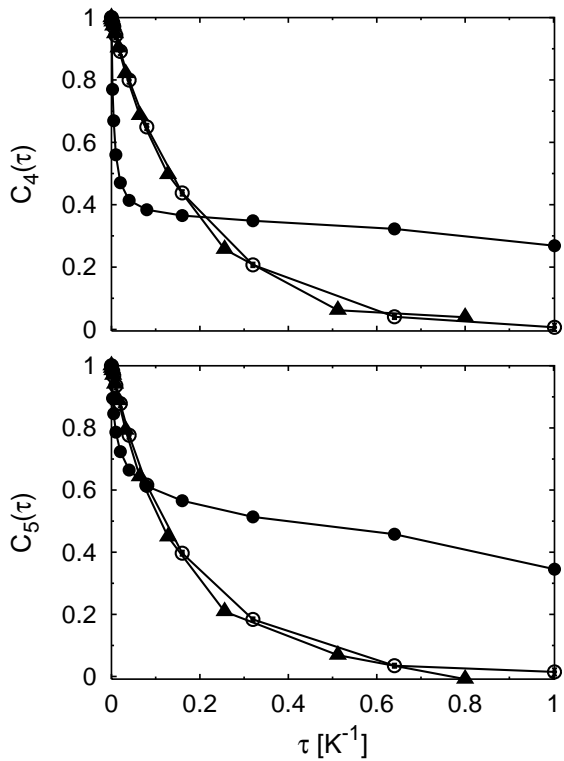}
    \hfill
  }
  \caption{
    Comparison of the multipole time-correlation functions 
    $c_L(\tau)$ for $L=4$ (upper panel) and $L=5$ (lower panel), 
    as calculated for the whole Li$^+$@He$_{70}$ (empty circles), within the 
    first solvent shell of the same cluster (filled circles), and for CO@He$_{15}$ 
    (triangles). 
     }
  \label{fig:multipoli-Li+He_70}
\end{figure}

In Fig.~\ref{fig:multipoli-Li+He_70} we compare the multipole time-correlation functions 
$c_L(\tau)$ for $L=4$ (upper panel) and $L=5$ (lower panel), 
as calculated for the first solvation shell in Li$^+$@He$_{70}$, for the same whole cluster, 
and for CO@He$_{15}$. 
The time correlations of the $L=4$ and $L=5$ multipoles of the first shell, 
albeit differing in magnitude, both show a very slow decay (almost a saturation in the case of $L=4$), 
whereas the same quantities, when considered for the whole cluster, 
decay rapidly to zero, very similarly to what happens in CO@He$_{15}$. Also 
the $L=6$ correlation function (not reported) exhibits a slow decay, just slightly more rapid than 
for the $L=4$ and $L=5$ curves. On the contrary, the decay of all the 
other $c_L(\tau)$ correlations, calculated within the first shell, is extremely fast.
We interpret this result as an indication that the $^4$He structure in the first 
solvation shell  
maintains an average shape, behaving like a solid cage which separates the Li$^+$ ion from  
the liquid environment of the outer shells.  
The signature of the rigidity of the first shell is lost among the random fluctuations 
in the more external regions, and thus it disappears in the multipole correlations 
of the entire system. 

At least qualitatively, we can attempt a better understanding of the behavior of the 
time-correlations $c_L(\tau)$.  
In a very simplified and ideal picture, the diffusion of the $^4$He atoms of a \emph{solid} first 
solvation shell during the simulation can be thought as the combination of two kind of motions: one 
approximatively corresponding to a coherent rotation of the whole shell, and the other 
due to high-frequency small-amplitude fluctuations about the smooth trajectory of the former. 
When the cluster configurations along the path are rotated backward in order 
to minimize the particles diffusion, the neat rotation is totally suppressed, 
whereas the fast fluctuations are not. Thus the only surviving motions can be seen as the 
incoherent fluctuations of $^4$He atoms in the first shell about average positions. 
More realistically, other motions, not removed by the backward rotation procedure, enter the 
decomposition of the $^4$He atoms diffusion in the first shell. 
If these motions are sufficiently slow, i.e. if their energies are sufficiently low, the average 
shape of the first shell is not disrupted too quickly. 
Correspondingly, since imaginary-time correlation functions can be written as a sum of decaying 
exponentials whose decaying constants are the excitation energies of the system,  
the $c_L(\tau)$ correlations, characteristic of the system geometry, exhibit a slow decay.

Within the same qualitative picture, the atoms in the outer regions undergo random motions, none of 
which is reducible to a neat rotation. The slowest modes of the whole 
cluster are thus fast enough to prevent the persistence of any global shape (and to produce 
a rapid decay of the corresponding time-correlations $c_L(\tau)$), but anyway they are much slower 
than the vibrations of the atoms in the first shell, thus explaining the lower decaying rate 
of the $c_L(\tau)$ correlation functions at small $\tau$ with respect to those of the 
first shell.

\begin{figure}[t]
  \hbox to \hsize{\hfill
    \includegraphics[width=75mm]{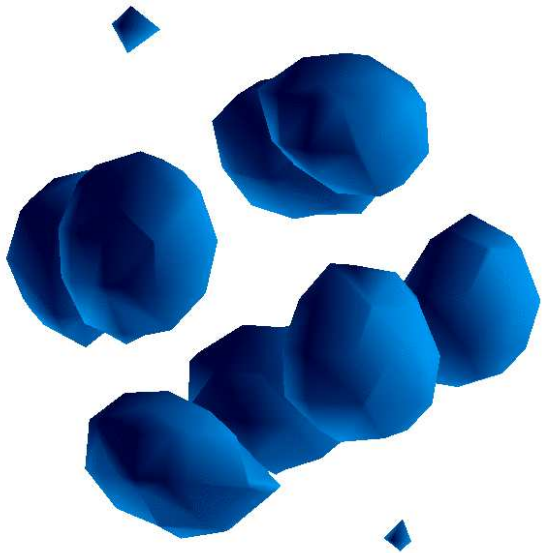}
    \hfill
  }
  \caption{
     He density distribution within the first solvation shell of 
     Li$^+$@He$_{70}$. 
     The displayed isosurface contains $60\%$ of the number of He atoms 
     in the first shell. The Li$^+$ ion is located at the center of the picture, 
     but it is not displayed.
     }
  \label{fig:rho-Li+He_70}
\end{figure}

Fig.~\ref{fig:rho-Li+He_70} depicts the $^4$He density distribution within the radius 
of the first shell, where we found an integrated density of $\sim 8.24$ $^4$He atoms. 
The solvent concentrates mainly in eight regions 
arranged in two parallel squares rotated by $\pi/4$ with respect to each other. 
At the opposite poles of this structure, along an axis perpendicular to the planes of 
the squares, two small lower-density accumulations, probably due to incursions 
of $^4$He density from the 
second shell, are found (they disappear well before the 
other eight ones when we increase the isodensity value chosen to draw the picture 
of Fig.~\ref{fig:rho-Li+He_70}). As we explained in discussing Li$^+$@He$_8$, 
for a two rotated-squares structure the lowest non-vanishing $q_L$ values correspond 
to $L=4$ and $L=5$. We note that the addiction of two more vertexes on a line perpendicular 
to the squares produces a polyhedron having the lowest non-vanishing $q_L$'s for $L=5$ and 
$L=6$. The behavior of the $c_L(\tau)$ correlation functions can thus be seen as 
the result of the concomitant signals corresponding to these two structures.
We stress at this point the need of considering dynamical correlations $c_L(\tau)$ of the 
multipoles of the cluster mass density, rather than their static magnitudes $q_L$, in order 
to reveal a solid-like behavior of the solvent. In fact, by calculating the quantities 
$q_L$ of the He density of the first solvation shell of Li$^+$@He$_{70}$, we obtain value at $L=4$ 
which is larger but of the same order of magnitude of the value at $L=3$, whose corresponding 
imaginary-time correlation function, instead, decays extremely rapidly.

\begin{figure}[t]
  \hbox to \hsize{\hfill
    \includegraphics[width=75mm]{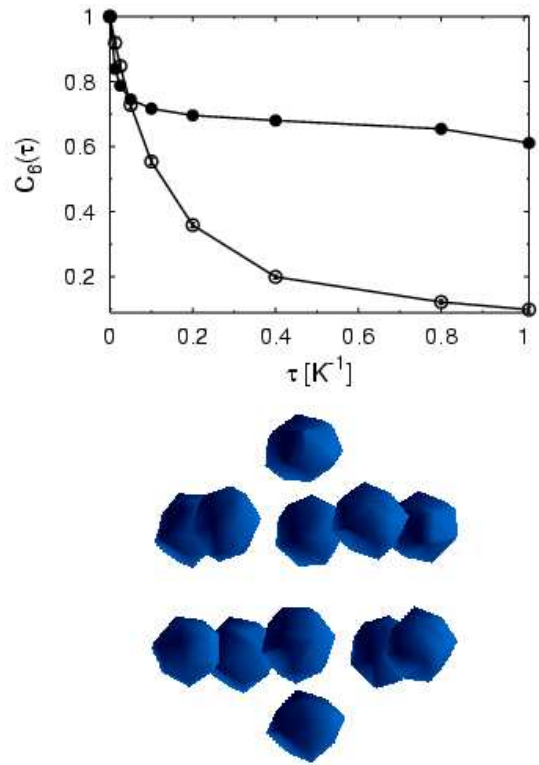}
    \hfill
  }
  \caption{
    Upper panel: multipole time-correlation functions $c_L(\tau)$ for $L=6$, as calculated for the 
    whole Na$^+$@He$_{70}$ cluster (empty circles) and within the 
    first solvent shell (filled circles). 
    Lower panel: He density distribution in the first shell of Na$^+$@He$_{70}$. The 
    isodensity surface contains the $60\%$ of the density of the 
    first shell. The Na$^+$ ion resides at the center of the structure and it is 
    not displayed. 
     }
  \label{fig:na+}
\end{figure}

Even more evident is the existence of a solid first shell in clusters 
doped with a Na cation. Our simulations indicate that the time 
correlations of the first five multipoles within the first shell decay 
rapidly, whereas the $L=6$ shows an extremely slow decay. These results 
are compatible with an icosahedral structure, whose lowest non-vanishing 
multipoles correspond to $L=6$ and $L=10$. For the first shell of Na$^+$@He$_{70}$, 
we show in Fig.~\ref{fig:na+} the slow-decaying $L=6$ multipole time-correlation 
function (upper panel) and the $^4$He density distribution (lower panel). 
The latter clearly shows an icosahedral order, in agreement with  previous QMC calculations, 
based on the shadow wave functions technique.~\cite{buzzacchi-2001}

Interestingly enough, despite the
stronger                            
interaction potential characterizing the Li$^+$-He pair,
we find a 
more pronounced solid-like behavior of the first shell of
$^4$He around a Na$^+$ ion than in the case of Li$^+$.
This is due to a subtle balance between the depth and the position of the 
minimum of the He-ion interaction potential, which determines the average number 
of $^4$He atoms in the first solvation shell, and, hence, the average distances  
among them. For instance, the nearest neighbor He atoms in the first shell 
of Li$^+$@He$_{70}$ are separated by roughly $2.6$ \AA, i.e. they find each other 
in the repulsive region of the He-He interaction potential. In contrast, the 
nearest neighbor distance in the first shell of Na$^+$@He$_{70}$ is 
slightly less than $2.8$ \AA, that is within the well of the 
He-He pair potential and closer to its minimum ($\sim 3.0$ \AA).   

The results for Na$^+$@He$_{70}$ have been obtained using a time step 
$\epsilon=\rm 0.0125~K^{-1}$. 
By integrating the $^4$He density within the first shell 
of Na$^+$@He$_{70}$, we obtain an estimate of 
twelve $^4$He atoms, in agreement with shadow wave function 
and path-integral Monte Carlo (PIMC) results,~\cite{buzzacchi-2001,rossi-thesis}
but at variance with other existing PIMC calculations.~\cite{na+nakayama}
Our result is stable, within the statistical error, when reducing the time step by a factor 
$2.5$, thus showing a negligible time-step dependence in this $\epsilon$ range.  
For larger time steps, however, the reliability of the simulations is not guaranteed. 
In Ref.~\onlinecite{na+nakayama}, where a time step of $\rm 1/20~K^{-1}$ is
used, sixteen $^4$He atoms are found, instead of twelve, in the first 
shell of a Na$^+$@He$_{100}$ cluster at the temperature of $\rm 1~K$.
The above discrepancy may originate from the choice of the He-Na$^+$ 
interaction potential~\cite{na+nakayama} or, more likely, 
it is a time-step effect. In fact, for strongly 
attractive potentials, such as the Li$^+$-He and the Na$^+$-He ones, 
we noted that the use of a large time step 
(like the one used in Ref.\onlinecite{na+nakayama})
may lead to overestimate the $^4$He density in the first shell. 
In particular, 
by simulating Na$^+$@He$_{70}$ with the same rather large time step $\epsilon=\rm~0.05 K^{-1}$,  
we reproduced the result of Ref.~\onlinecite{na+nakayama}. We also  
observed that the $^4$He density in the first solvation shell is much more delocalized, 
lacking the icosahedral structure previously found, and that the lowest twelve 
multipoles time-correlation functions decay now much more rapidly.

\begin{figure}[t]
  \hbox to \hsize{\hfill
    \includegraphics[width=75mm]{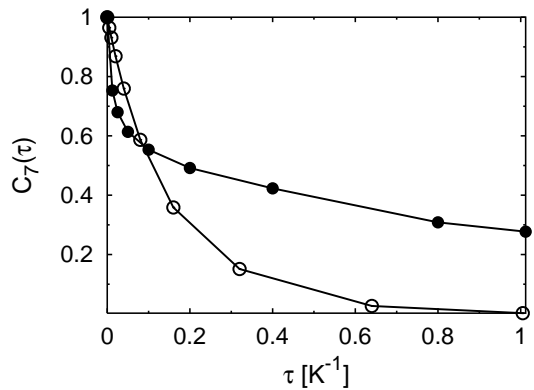}
    \hfill
  }
  \caption{
    Multipole time-correlation functions $c_L(\tau)$ for $L=7$, as calculated for the 
    whole Be$^+$@He$_{70}$ cluster (empty circles) and within the 
    first solvation shell (filled circles). 
     }
  \label{fig:be+multipoli}
\end{figure}

\subsubsection{Alkaline-earth ions doped $^4$He clusters}

We now turn our attention to the
case of $^4$He clusters doped with alkaline-earth ions.
Due to their larger radii and their relatively weaker interactions with helium atoms, 
these ions are expected to give rise to bubble-like structures of the 
$^4$He liquid around the impurity. 
This expectation is confirmed by mobility measurements at finite 
temperature for several alkaline-earth ions but not for Be$^+$, for 
which results of similar measurements suggest instead a snowball 
structure of the defect.~\cite{foerste-1997} 
For this reason, in the following, we will focus particularly 
on the solvent structure around Be$^+$. 
The rigidity of the defect structures of Be$^+$ and Mg$^+$ can hardly be 
distinguished on the basis of static correlation 
functions among $^4$He atoms,~\cite{rossi-2004} thus spurring us 
to consider the multipole moments dynamical correlations 
criterion.~\cite{baroni-moroni-chemphyschem}

\begin{figure}[t]
  \hbox to \hsize{\hfill
    \includegraphics[width=70mm]{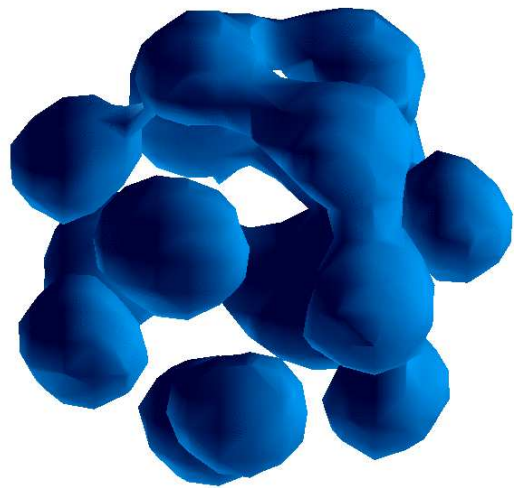}
    \hfill
  }
  \caption{
    He density distribution in the first shell of Be$^+$@He$_{70}$. The 
    isodensity surface contains the $60\%$ of the density of the 
    first shell. The Be$^+$ ion resides at the center of the structure and it is 
    not displayed. 
     }
  \label{fig:be+rho}
\end{figure}

In order to detect even weak signatures of a possible solid-like order 
around the alkaline-earth ions, the clusters configurations, used for 
calculating both static and dynamical quantities, have been rotated back  
in order to minimize the particles diffusion within the 
\emph{first} solvation shell, and not in the whole clusters. 
According to our results, all the multipoles correlations  of the 
first shell of Be$^+$@He$_{70}$ up to $L=5$ decay rapidly (even faster than in the whole cluster), 
whereas for $L=6$ and, more evidently, for $L=7$ the correlation 
decay is much slower than that of the whole cluster. Also in this case 
the magnitude of the multipoles $q_L$ provide a guide in the search of 
the slow decaying time-correlations.  
Fig.~\ref{fig:be+multipoli} displays the $L=7$ 
multipole correlation in the first solvation shell and in the whole cluster. 
In spite of the faster decay than for Li$^+$@He$_{70}$ and Na$^+$@He$_{70}$, 
the behavior of $c_L(\tau)$, for $L=7$, indicates an extremely slow disappearance of 
the shape assumed by the solvent around the Be$^+$ ion. The average shape of the solvent in the first shell is 
shown in Fig.~\ref{fig:be+rho}. Despite being less localized than 
for alkali ion doped clusters, the $^4$He density in the vicinity of Be$^+$ appears 
anyway significantly structured.  

\begin{figure}[t]
  \hbox to \hsize{\hfill
    \includegraphics[width=75mm]{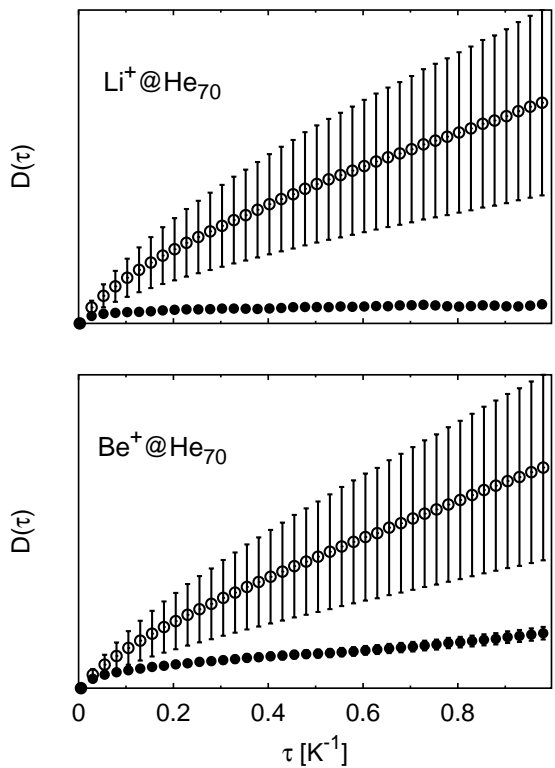}
    \hfill
  }
  \caption{
    Upper panel: diffusion coefficient $D(\tau)$ as a function of (imaginary) time 
    $\tau$ calculated from the backward-rotated configurations, 
    for the He atoms in the whole 
    Li$^+$@He$_{70}$ cluster (empty circles), and for the He atoms in its first solvation shell 
    (filled circles). 
    Lower panel: same quantities for Be$^+$@He$_{70}$, displayed with the same symbols.
     }
  \label{fig:diffusion-be-li}
\end{figure}

Fig.~\ref{fig:diffusion-be-li} compares the diffusion coefficient in imaginary time, $D(\tau)$, as calculated 
after the backward rotation, 
for the $^4$He atom in the whole cluster and within the first shell, for Li$^+$@He$_{70}$ and Be$^+$@He$_{70}$. 
For both systems, with increasing the (imaginary) time $\tau$, the coefficient $D(\tau)$ of the whole cluster 
increases much more than within the first solvation shell, due to the liquid character of the external regions.  
We interpret the small diffusion rate of the He atoms in the first shell, and, moreover, the close similarities 
in $D(\tau)$ for Be$^+$@He$_{70}$ and Li$^+$@He$_{70}$ (which we consider a prototype of a cluster 
with a first shell with solid-like order) as an indication of a nearly rigid-body behavior  
of the first shell of Be$^+$@He$_{70}$. This conclusion will be strengthened by the comparison with the results 
for Mg$^+$@He$_{70}$, which will be discussed next.

\begin{figure}[b]
  \hbox to \hsize{\hfill
    \includegraphics[width=75mm]{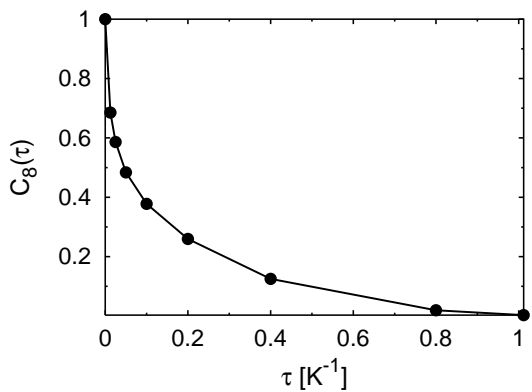}
    \hfill
  }
  \caption{
    Multipole time-correlation function $c_L(\tau)$ for $L=8$, as calculated within 
    the first solvation shell of Mg$^+$@He$_{70}$. 
     }
  \label{fig:mg+multipoli}
\end{figure}

For Mg$^+$@He$_{70}$ the calculation of $q_L$, within the first solvation shell, gives a 
faint signal for $L=8$. However, the corresponding imaginary-time correlation function 
$c_L(\tau)$, which is reported in Fig.~\ref{fig:mg+multipoli}, exhibits an extremely fast decay, 
despite being the slowest decaying one among those we have analyzed (up to $L=12$). 
Our findings suggest that the He atoms surrounding Mg$^+$  undergo random fluctuations 
which disrupt the local structure of the solvent much more quickly than in the other cases 
examined so far, thus preventing the formation of any longly persisting solid-like order. 
For Mg$^+$@He$_{70}$ the rapid decaying of the $c_L(\tau)$ correlations of the first shell 
is very similar to that of the corresponding ones of the entire cluster, thus indicating 
a liquid-like behavior of the whole system. 

\begin{figure}[t]
  \hbox to \hsize{\hfill
    \includegraphics[width=75mm]{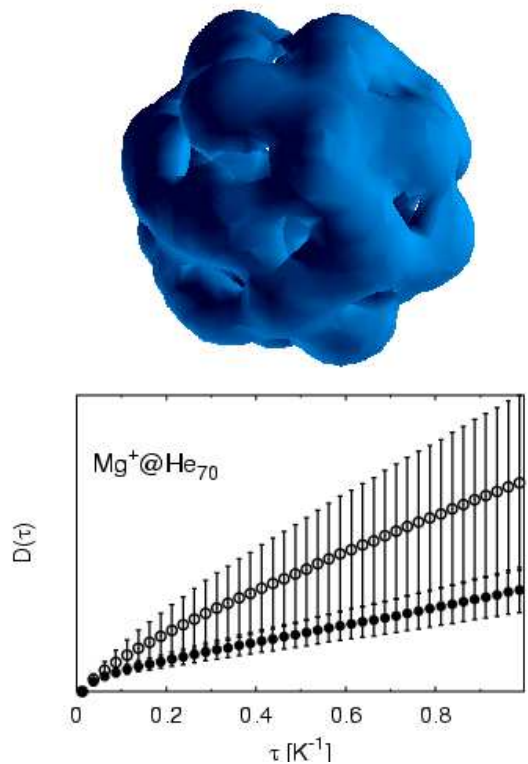}
    \hfill
  }
  \caption{
    Upper panel: He density distribution in the first shell of Mg$^+$@He$_{70}$. 
    The integral of the density contained within the displayed isodensity surface 
    amounts to $60\%$ of the number of He atoms in the first shell. 
    Lower panel: diffusion coefficient $D(\tau)$ as a function of (imaginary) time 
    $\tau$ for the He atoms in Mg$^+$@He$_{70}$, as calculated form the backward-rotated configurations, 
    for the whole cluster (empty circles) and for the first solvation shell (filled circles). 
     }
  \label{fig:mg-rho-diffusione}
\end{figure}

As it is shown in the upper panel of Fig.~\ref{fig:mg-rho-diffusione}, the $^4$He 
density distribution in the first shell is considerably more diffuse than what 
found for the other ions. 
By increasing the isodensity value chosen for the graph, it becomes possible to 
identify regions of higher density, nevertheless the solvent is much more delocalized 
than for Be$^+$@He$_{70}$. 
The lower panel of Fig.~\ref{fig:mg-rho-diffusione} compares the imaginary-time diffusion 
of the $^4$He atoms in the entire cluster and in the first solvation shell, after the backward rotation. 
The two curves are 
now closer and they even merge within the error bars, thus proving that the rotational 
diffusion of the particles in the first shell cannot be reduced to a neat rotation. 
The liquid-like behavior of the solvent is even more evident in Ca$^+$@He$_{70}$, for which 
the first shell multipole time-correlation functions decay very rapidly, the density distribution 
in the first solvation shell is largely delocalized and the diffusion coefficients 
of $^4$He atoms of the inner and of the outer shells are much less distinguishable.

\section{Conclusions}

Using PIGS simulations, we have studied the solvation 
of alkali and alkaline-earth ions in $^4$He clusters. In order to
discriminate between solid-like versus liquid-like 
behavior of the solvent in the vicinity of the impurity, 
we employed a criterion based on dynamical correlations 
of the multipole moments of the $^4$He density in the first shell, 
previously proposed for quantum clusters consisting of just one completed 
shell. We also used other indicators to clarify the structure of $^4$He
around the impurity.

The dynamical correlations criterion, whose applicability to clusters 
with more than one shell was not granted a priori, proved suitable to address  
situations in which regions with different rigidity coexist within the system, without 
requiring any prior knowledge of the defect structure. 
In particular we find bubble (liquid-like) structures for heavier alkaline-earth 
ions (Mg$^+$ and Ca$^+$), whereas alkali-ions are caged by a solid-like, snowball
structure, further embedded in an outer liquid environment. For these two categories, 
whose features are expected to be different both theoretically~\cite{bachman,flavio-1978} and 
experimentally,~\cite{glaberson-1975,foerste-1997} the 
multipole correlations signatures are 
clearly different and unambiguously distinguishable. 
 
For Be$^+$, our simulations indicate a border-line behavior, 
which, however, has much more similarities with alkali ions 
than with heavier alkaline-earth ones. This results suggests 
that Be$^+$ forms a snowball upon solvation, possibly elucidating  
the ``anomalous'' mobility of Be$^+$ in liquid helium 
with respect to Mg$^+$, Ca$^+$ and other alkaline-earth ions.~\cite{foerste-1997} 

\vbox to 0pt{\vss}
%
%

\end{document}